\documentstyle[aps,prl,twocolumn,epsf]{revtex}

\renewcommand{\v}[1]{{\bbox #1}}
\newcommand{\Eq}[1]{Eq.~(\ref{#1})}
\newcommand{\vev}[1]{\left\langle #1 \right\rangle}
\newcommand{\vphi}{\varphi}
\newcommand{\la}{\lambda}

\begin{document}
\draft
\wideabs{
\title{Binding Transition in Quantum Hall Edge States}
\author{Hsien-chung Kao$^a$, Chia-Hung Chang$^b$, and Xiao-Gang Wen$^{c,b}$}
\address{a) Department of Physics, Tamkang University, Tamsui,
Taiwan 25137, R.O.C.}
\address{
b) Physics Division,
National Center for Theoretical Sciences,
P.O.Box 2-131, Hsinchu, Taiwan 300, R.O.C.}
\address{c) Department of Physics, Massachusetts Institute of Technology,
Cambridge, MA 02139}
\date{July 1, 1998}
\maketitle
\begin{abstract}
We study a class of Abelian quantum Hall (QH) states which are topologically
unstable (T-unstable).
We find that the
T-unstable QH states can have a phase transition on the edge which
causes a binding between electrons and reduces the number of gapless edge
branches.  After the binding transition, the single-electron
tunneling into the edge gains a finite energy gap, and only certain
multi-electron co-tunneling
(such as three-electron co-tunneling for $\nu=9/5$ edges)
can be gapless.
Similar phenomenon also appear for edge state on the boundary between
certain QH states. For example edge on the boundary between
$\nu=2$ and $\nu=1/5$ states only allow three-electron co-tunneling
at low energies after the binding transition.
\end{abstract}


}


A large class of quantum Hall (QH) liquids \cite{QH1}
(almost all those observed in
experiments) is called Abelian QH state. The topological orders
\cite{top,rev}
in the
Abelian QH liquids are labeled by a rank $\kappa$ integer symmetric matrix
$K$
(called the $K$-matrix) and a $\kappa$ dimensional integer vector $\v t$
(called the charge vector).\cite{Km}
The effective theory is given by the $U(1)$ Chern-Simons (CS) theory:
\begin{equation}
L_{bulk} ={1\over 4\pi} K_{ij} a_{i\mu}\partial_\nu
a_{j\lambda}\epsilon^{\mu\nu\lambda}
-eA_\mu t_i \partial_\nu a_{i\lambda}\epsilon^{\mu\nu\lambda}.
\label{Lbulk}
\end{equation}
The above pure CS effective theory is dual to the Ginzburg-Landau-CS
effective theory obtained earlier.\cite{GL}
$(K, \v t)$ determine all the universal properties of the Abelian QH states.
For example,
all the allowed quasi-particles are labeled by $\kappa$ dimensional non-zero
vector ${\v l}$ with integer
elements.  The electric charge and statistics of
a quasi-particle are given by
\begin{equation}
\theta=\pi {\v l}^{\rm T} K^{-1} {\v l}, \ \ \
Q_q=-e{\v t}^{\rm T} K^{-1} {\v l},
\end{equation}
while the filling fraction is given by
$
\nu = {\v t}^{\rm T} K^{-1} {\v t}.
$.
$(K, \v t)$ also determines the structure of edge excitations\cite{rev}
(at least for the sharp edges \cite{CW}).

According to Haldane,\cite{Htopins}
 an Abelian quantum Hall theory is T-unstable if there
exist quasi-particles (labeled by ${\v m}$),
that are both bosonic and charge neutral, i.e.
\begin{equation}
{\v m}^{\rm T} K^{-1} {\v m}=0;\ \ \ {\v t}^{\rm T} K^{-1} {\v m}=0.
\label{null}
\end{equation}
Such a vector $\v m$ will be called neutral null vector.
Since these quasi-particles carry trivial quantum numbers, their creation
and annihilation operators can appear in the Hamiltonian without breaking
any
symmetries.  By including them in Eq. (\ref{Lbulk}),  a more generic
effective
theory can be obtained.
In this paper we will study the physical consequences of the these operators
on the edge states. We find that, under right conditions,
they can cause a phase transition in the edge states.
Such a transition will be called binding transition in this paper.

Many QH states have  neutral null vectors, such as the $\nu=9/5$ state with
\begin{equation}
K=\left(\matrix{
1&0&0\cr
0&1&0\cr
0&0&-5}\right), \ \ \
 {\v t}=\left(\matrix{
1\cr
1\cr
1}\right) .
\label{Kt}
\end{equation}
Before the binding transition, the edge state of the above QH state
has three branches and the single-electron tunneling is gapless.
After the binding transition, as we will see,
the edge state has only one gapless branch, and the single-electron
tunneling
opens up a
finite energy gap.
Only three-electron co-tunneling is gapless.

The binding transition can also appear on the boundary
between two different QH liquids.
The edge state between two different QH liquids $(K_1, \v t_1)$ and
$(K_2, \v t_2)$ is described by $K=K_1\oplus (-K_2)$,
$\v t=\v t_1 \oplus \v t_2$.
The edge state will be called  T-unstable if the
$(K, \v t)$ has
non-zero neutral null vector $\v m$.
For a sequence of hierarchical QH liquids $\nu=1/3,2/5,3/7,...$.
the edge
state between any two QH states in the sequence is T-unstable since they are
based
on the same $\nu=1/3$ state. The edge excitations from the base
1/3 state can annihilate each other.
More non-trivial cases of T-unstable boundary
can appear between hierarchical
states based on different QH liquids.
We find T-unstable edge states between
$(2/5, 2/9)$, $(2/5, 3/13)$, $(2/9,2/7)$,
$(2/5, 1/7)$, $(2/5, 2/13)$, $(1/9, 2/9)$,... QH states.
The simplest T-unstable edge state is the one between
$\nu=2$, and $\nu=1/5$ states. Such a edge state is equivalent to that
of the $\nu=9/5$ state and have the same binding
transition.



To show the above results,
let us concentrate on the simplest T-unstable systems,
the $\nu=9/5$ state described by \Eq{Kt}.
The quasi-particle operators are labeled by integer vectors
$\v l$. The electron operators are those of the quasi-particle operators
which
carry charge $e$ and have statistics $\theta=(2n+1)\pi $ ({\it ie} the Fermi
statistics). There are infinite many electron operators, which can be
labeled
by two integers $(k_1,k_2)$:
\begin{equation}
 \v l_e=\pmatrix{ k_1\cr k_2\cr -5(1-k_1-k_2)\cr}
 \label{l-e}
\end{equation}
The $\nu=9/5$ state has infinite many neutral null vectors:
integer$\times \v m_1$ and integer$\times \v m_2$, where
\begin{equation}
{\v m}_1 = \left(\matrix{
-1\cr
2\cr
5}\right),
{\v m}_2 = \left(\matrix{
2\cr
-1\cr
5}\right)
\label{m12}
\end{equation}
It is known that edges of quantum Hall systems are described by a
chiral Luttinger liquid ($\chi$LL) theory.  In imaginary time, the
corresponding $\chi$LL action contains $N$ bosonic fields $\phi_i$ and has
the
form~\cite{rev}
\begin{equation}
S_{edge} = {1 \over 4 \pi} \int{dx \, d\tau\, [iK_{ij} \partial_x \phi_i
\partial_\tau \phi_j +
V_{ij} \partial_x \phi_i \partial_x \phi_j]}.
\label{sec:action}
\end{equation}
%
On the edge, quasi-particles are created by the vertex operators, $V_{\v l}
= \exp(i l_j \phi_j)$.
The correlation function of $V_{\v l}$ have a form
\begin{eqnarray}
(\prod_{k=1}^{N^+} (x + i v^+_k \tau)^{-\alpha_k}) (\prod_{k=1}^{N^-}
(x - i v^-_k \tau)^{-\beta_k}).
\label{sec:exponents}
\end{eqnarray}
Here $N^+$ and $N^-$ are the numbers of positive and negative eigenvalues of
$K$; $v^\pm_k, \alpha_k, \beta_k$ are nonnegative real numbers which depend
on
$V$ and $K$.
Setting all velocities $v^{\pm}_k =
v$ and introducing $z = x + i v \tau$, we have
\begin{equation}
\langle e^{i l_j \phi_j(x,\tau)} e^{-i l_j \phi_j(0,0)} \rangle \propto
{1 \over z^{K({\v l})}} {1 \over |z|^{2 \Delta({\v l}) - K({\v l})}}.
\end{equation}
Here
$\Delta({\v l}) \equiv
(\sum_{k=1}^{N^+} \alpha_k + \sum_{k=1}^{N^-} \beta_k)/2$ is the scaling
dimension of the operator $\exp(i l_j \phi_j)$, and
$K({\v l}) \equiv l_i (K^{-1})_{ij} l_j = \sum_{k=1}^{N^+} \alpha_k -
\sum_{k=1}^{N^-} \beta_k$,
which is independent of $V$ and determines the statistics of the
quasiparticle.
Since the scaling dimension of an operator is a
functions of $V$, it is  useful to write $V$ in such a way that isolates the
parts of $V$ affecting $\Delta({\v l})$. We will follow the approach used in
Ref. \cite{MW}.

Eq. (\ref{sec:exponents}) is obtained by simultaneously diagonalizeing $K$
and $V$ through a basis change $\phi_i = M_{ij}\tilde \phi_j$.  This can be
done in
two steps. First, we find an $M_1$ that brings $K$ to the pseudo-identity
$I_{N^{-},N^{+}}$, i.e.,
\begin{equation}
M_1^{\rm T} K M_1 = I_{N^{-},N^{+}} = \left(\matrix{
-I_{N^{-}}&0\cr
0&I_{N^{+}}}\right).
\end{equation}
For the state $\nu=9/5$, we find
\begin{equation}
M_1=\left(\matrix{
{-5\lambda^2\over 2}&{-2+5\lambda^2\over 2}&{-5\lambda\over \sqrt{5}}\cr
2+5\lambda+5\lambda^2&-5\lambda-5\lambda^2&{5+10\lambda\over \sqrt{5}}\cr
{-2-4\lambda-5\lambda^2\over 2}&{4\lambda+5\lambda^2\over
2}&{-2-5\lambda\over
\sqrt{5}}}\right),
\label{M1I}
\end{equation}
and $(N^-,N^+)=(1,2)$.
Then we find the second basis change $M_2$ that diagonalizes $V$ while
leaves
the pseudo-identity invariant.  In other words, $M_2$ belongs to the proper
pseudo-orthogonal group $SO(N^{-},N^{+})$.
In the new basis $\tilde{\v \phi} = (M_1 M_2)^{-1} {\v \phi}$,
\begin{eqnarray}
\label{sec:eqf}
\tilde{\v l}&=& M_2^{\rm T} M_1^{\rm T} {\v l}\nonumber \\
\tilde K=I_{N^{-},N^{+}} &=& M_2^{\rm T} M_1^{\rm T} K M_1 M_2 \\
\tilde V &=& M_2^{\rm T} M_1^{\rm T} V M_1 M_2. \nonumber
\end{eqnarray}
Thus the functions $K({\v l})$ and $\Delta({\v l})$ are basis-independent.
Now that both $\tilde V$ and $\tilde K^{-1}$ become diagonal,
the correlation functions are
trivial:
$
\langle e^{i \tilde{\phi}_j(x,\tau)} e^{- i \tilde{\phi}_j(0,0)} \rangle
\propto
{1 \over x \mp i v_j \tau}
$
where the sign depends on whether $\tilde \phi_j$ appears with $-1$ or $+1$
in
$I_{N^{-},N^{+}}$.  Consequently,
the scaling dimension of the operator $\exp(i
l_j \phi_j)$ is found to be
$
\Delta({\v l})=\tilde{l}_j \tilde{l}_j = l_i \Delta_{ij} l_j,
$
with
\begin{equation}
2 \Delta = M_1 M_2 M_2^{\rm T} M_1^{\rm T}.
\label{sec:eq1}
\end{equation}
Drawing analogy from
special relativity, we can factor $M_2$ into a product of a
symmetric positive matrix $B$ analogous to the Lorentz boost and an
orthogonal
matrix $R$: $M_2=BR$.  Eq. (\ref{sec:eq1}) becomes
\begin{equation}
2 \Delta = M_1 M_2 M_2^{\rm T} M_1^{\rm T}
= M_1 B^2 M_1^{\rm T}.
\end{equation}

For the $\nu=9/5$ state, the matrices $B$ and $R$
can be parameterized as
\begin{eqnarray}
B&=&\left(\matrix{
\gamma &\gamma\beta_1 &\gamma\beta_2 \cr
\gamma\beta_1 &1+{\gamma^2\beta_1^2 \over \gamma+1} &{\gamma^2\beta_1\beta_2
\over \gamma+1} \cr
\gamma\beta_1 &{\gamma^2\beta_1\beta_2 \over \gamma+1} &1+{\gamma^2\beta_2^2
\over \gamma+1} }\right),\nonumber\\
R&=&\left(\matrix{
1 &0 &0 \cr
0 &\cos\vphi &-\sin\vphi \cr
0 &\sin\vphi &\cos\vphi } \right)
\label{sec:boostdef}
\end{eqnarray}
where $\gamma=1/\sqrt{1-\beta^2}$.
Note that
$V= (M_1^T)^{-1} (M_2^T)^{-1} \tilde V (M_1)^{-1}(M_1)^{-1}$, thus
$(\lambda,\beta_1,\beta_2,\vphi)$ and the three diagonal elements in
$\tilde V $, $(v_1,v_2,v_3)$,
can be viewed as a parameterization of $V$. Since $V$ contains only six
independent parameters, we may
set one of the above seven parameters to zero.


For neutral null vectors $\v m_{1,2}$, the quasi-particle
operators
$V_{\v m_1} \equiv e^{i  p_1x} \exp\left\{i(\v m_1)_j \phi_j\right\}$
and
$V_{\v m_2} \equiv e^{i  p_2x} \exp\left\{i(\v m_2)_j \phi_j\right\}$
carry trivial quantum number \cite{comm1} and the generic edge
Hamiltonian/action contains
a term $ \Gamma_1 V_{\v m_{1}} + \Gamma_2 V_{\v m_{2}} +h.c. $.
The problem here is that how
this quasi-particle
term affect the dynamics of low lying edge excitations.
First let us consider when the quasi-particle
term becomes a relevant perturbation.

The scaling dimensions of $ V_{\v m_{1,2}}$ are found to be
$
\Delta({\v m}_1) =  \gamma^2\left\{-1+\beta_1\right\}^2
$
and
$
\Delta({\v m}_2) ={1\over 4} \gamma^2
\big\{-(14 + 30\lambda + 45\lambda^2) +
(-4 + 30\lambda + 45\lambda^2)\beta_1 -6\sqrt{5}(1 +
3\lambda)\beta_2\big\}^2
$.
Note that $\Delta(\v m_{1,2})$ depend only on $(\lambda, \beta_1,\beta_2)$,
and in the rest of the paper
we will set one of the redundant parameter $\beta_2=0$.

In one area of the
$\beta_1$-$\lambda$ plan, we find that
both $\Delta(\v m_{1,2}) >2$ (see Fig. 1). In this case, both
the neutral quasi-particle operators are irrelevant and can be dropped at
low
energies. Therefore in this area of the
$\beta_1$-$\lambda$ plan, the  neutral quasi-particle operators  do not
cause
any instability, and the edge theory is still described by
Eq. (\ref{sec:action}), and has three gapless branches.
We also see in Fig. 1 that in another area, $\Delta(\v m_{1}) <2$
and $\Delta(\v m_{2}) >2$. In this case only the $V_{\v m_1}$
is relevant. In the following, we will drop the $V_{\v m_2}$ term and
consider the effect of the $V_{\v m_1}$ term.
\begin{figure}
\epsfysize=2truein
\centerline{\epsffile{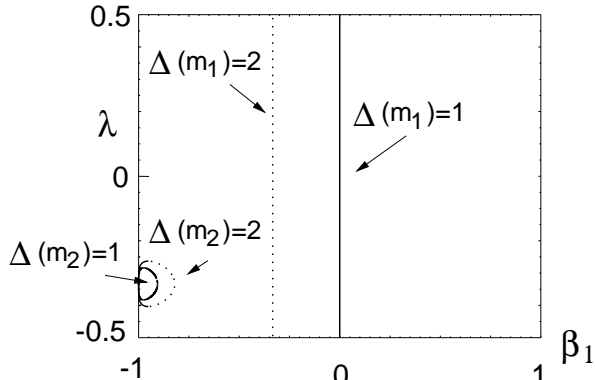}}
\caption{Plot of scaling dimension of the two neutral null operators for
$\nu = 9/5$ edge as functions of $(\beta_1, \lambda)$.  Dash lines indicate
when
operators become marginal ($\Delta({\v m}) = 2$) and solid lines indicate
the soluble case where ($\Delta({\v m}) = 1$).}
\label{fig1}
\end{figure}

In general, it is very difficult to solve our model with a relevant
quasi-particle operator. However, in a special case when $(\tilde \v
m_1)_3=0$
(where $\tilde \v m_1 = (M_1M_2)^T \v m_1$), the
problem can be simplified. This is because when $(\tilde \v m_1)_3=0$ the
mode described by $\tilde \phi_3$ is decoupled from the modes described by
$\tilde \phi_{1,2}$ even in the presence of the $V_{\v m_1}$ quasi-particle
operator. In this case, at least the sector described by $\tilde \phi_3$ can
be
solved, which describes a gapless edge mode.
The condition for  $(\tilde \v m_1)_3=0$ can be satisfied if and only if
$
\vphi=0,\pi
$.

Now let us concentrate on the dynamics of the decoupled sector described by
$\tilde \phi_{1,2}$. First we will show that if in addition to
$\vphi=0,\pi$, we
also have $\Delta(\v m_1)=1$, then the $\tilde \phi_{1,2}$
sector can be solved exactly.
The condition for the operator $V_{\v m_1}$
to have a scaling dimension 1 is
$ \beta_1=0 $.
Under this condition (note we have already set $\beta_2=0$ and
 $\vphi=0,\pi$), we have $\tilde m_1=(-1,1,0)$.
We can fermionize
$\exp(i\tilde \phi_1) \propto  e^{\alpha_1 x} \psi_1$, $\exp(i\tilde \phi_2)
\propto
e^{i\alpha_2 x} \psi_2$ with properly chosen $\alpha_{1,2}$ to
obtain $V_{\v m_1}\propto \psi_1^\dag\psi_2$ and
\begin{eqnarray}
{1 \over 4 \pi} & [ &
\sum_{a=1,2} \psi^\dagger_a(\partial_t + (-)^a v_a \partial_x+\mu)\psi_a
+ (\Gamma\psi^\dagger_1 \psi_2 + h.c.) \nonumber\\
&\;&+ \partial_x \tilde{\phi}_3 \partial_\tau \tilde{\phi}_3
+ v_3 (\partial_x \tilde{\phi}_3)^2].
\label{sec:action3}
\end{eqnarray}
with $\mu= p_1 (v_1+v_2)$.
Now we see that the system  described by Eq. (\ref{sec:action3})
is a free fermion theory and is exactly soluble.
We find that when $\Gamma < \mu$ the $\psi_{1,2}$ sector is gapless.
The low lying excitations are described by gapless free fermions
$\psi_{1,2}$ and bosons $\tilde \phi_3$, and form three gapless branches.
We will call such a phase three-branch phase.
When $\Gamma > \mu$, the $\psi_{1,2}$ sector
has a finite energy gap, and there is only one gapless branch described by
$\tilde \phi_3$. Such a phase will be called one-branch phase.

We note that the one-branch
phase is very stable. Respect to the one-branch fixed point, any change in
$V$ (or in
$\la,\beta_1,\vphi,v_1,v_2, v_3$) corresponds to an irrelevant or exactly
marginal perturbation.
(Only a change in $v_3$ corresponds to an exactly marginal perturbation.)
In particular if we change $\vphi$ away form $0,\pi$, it will flow back
to $0,\pi$ at low energies.
Therefore, the edge can be in the one-branch phase for a finite volume
in the parameter space of $V$ if $\Gamma$ is large enough.
The gapless branch is always described by $\tilde \phi_3$.
$\vphi$ will flow to $0$ or $\pi$ in the one-branch phase.

Now let us consider the correlation of a generic quasi-particle operator
$V_{\v l}\equiv e^{i\v l \cdot \v \phi}= e^{i\tilde \v l \cdot \tilde \v
\phi}$ where $\tilde \v l = M_2^T M_1^T \v l$.
(Note that, when $\v l=\v l_e$, $V_{\v l}$ will
describe an electron operator.)
In the three-branch phase, $V_{\v l}$
has an algebraic correlation which can be calculated through the
bosonization.  In the one-branch phase,
$\vphi$ flows to $0,\pi$, $(\tilde {\v m}_1)_3=0$
and the $\tilde \phi_3$ sector decouples with the $\tilde \phi_{1,2}$
sector.
We may write $V_{\v l}$ as
$
 V_{\v l}=V^\prime_{\v l}e^{i(\tilde {\v l})_3 \tilde \phi_3}$ and
$V^\prime_{\v l}= e^{i(\tilde {\v l})_1 \tilde \phi_1+i
 (\tilde {\v l})_1 \tilde \phi_2}
$.
It is easy to find
$\vev{ e^{i(\tilde {\v l})_3 \tilde \phi_3} e^{-i(\tilde {\v l})_3 \tilde
\phi_3}} = (x-vt)^{- (\tilde {\v l})_3^2}$.
Thus we can concentrate on the $\tilde \phi_{1,2}$ sector and the
correlation
of $V^\prime_{\v l}$ .

First we note that if we write the edge partition function in the form of
imaginary-time path integral and expand it in power of
$\Gamma$, the edge partition function will have the same form as the
partition
function of a 2D Coulomb gas. The ``particles'' in the Coulomb gas
correspond
to $V_{\v m_1}$ and $V^\dag_{\v m_1}$. The interaction potential
is given by
\begin{equation}
 -\ln \vev{ V_{\v m_1}(z) V^\dag_{\v m_1}(0)}
= \Delta( {\v m}_1 ) \ln |z|^2
\end{equation}
if we assume, for the time being, $ p_1=0$.  When
$\Delta( {\v m}_1 )<2$, the Coulomb gas is in the plasma phase and the
$\tilde \phi_{1,2}$ sector has a finite energy gap.
In the Coulomb gas picture, calculating the correlation of $V^\prime_{\v l}$
corresponds to calculating the change $\Delta E$
in the energy of the Coulomb gas when
we insert two test charges correspond to $V^\prime_{\v l}$ and
$V^{\prime\dag}_{\v l}$. (The correlation function is given by $e^{-\Delta
E}$.)
{}From the correlation
\begin{equation}
 -\ln \vev{ V^\prime_{\v l}(z) V^\dag_{\v m_1}(0)...}
= (\tilde {\v l})_1 (\tilde {\v m}_1)_1
\ln z^* + (\tilde {\v l})_2 (\tilde {\v m}_1)_2 \ln z + ...
\end{equation}
we see that if
\begin{equation}
-(\tilde {\v l})_1 (\tilde {\v m}_1)_1
+(\tilde {\v l})_2 (\tilde {\v m}_1)_2 =
\tilde {\v l} \tilde K^{-1} \tilde {\v m}_1 =\v l K^{-1} \v m_1=0,
\end{equation}
then $V^\prime_{\v l}$ will indeed correspond to a charged particle in the
Coulomb gas, since the interaction potential is real. Thus in the plasma
phase,
$V^\prime_{\v l}$ will have a finite and constant correlation at long
distance
due to the complete screening of the  plasma phase which gives $\Delta E=0$.
Now $V^\prime_{\v l}$
can be replaced by a pure number and the correlation of
$V_{\v l}$ is just $(x-vt)^{- (\tilde {\v l})_3^2}$.
Note that $ (\tilde {\v m}_1)_1^2=  (\tilde {\v m}_1)_2^2$ and
$\tilde {\v l} \tilde K^{-1} \tilde {\v m}_1=0$ requires
$(\tilde {\v l})_1^2=  (\tilde {\v l})_2^2$, thus
$ (\tilde {\v l})_3^2 =-(\tilde {\v l})_1^2+ (\tilde {\v l})_2^2 +
(\tilde {\v l})_3^2 = \tilde {\v l} \tilde K^{-1} \tilde {\v l} =
\v l K^{-1} \v l$. Therefore,
$
 \vev{V_{\v l} V_{\v l}^\dag} = (x-vt)^{ - \v l K^{-1} \v l}
$
in the plasma phase when $\v l K^{-1} \v m_1 =0$.

When $\v l K^{-1} \v m_1 \neq 0$, the interaction potential is
a complex function.\cite{Km}
As the particles in the Coulomb gas move around the test charge, the
partition function can have arbitrary phases which average out to zero,
unless the two test charges sit
at the same space-time point. Therefore, we expect $V^\prime_{\v l}$ to
have a short ranged correlation in the plasma phase. As a consequence
$V_{\v l}$ also has a short ranged correlation in the plasma phase
when $\v l K^{-1} \v m_1 \neq 0$.

In the above we have assumed that $ p_1=0$. If $ p_1\neq 0$ then
we need $\Gamma >\mu$ to open an energy gap in the $\tilde \phi_{1,2}$
sector
and to be in the one-branch phase.
All the above results remain to be valid if we regard the plasma phase
mentioned above as the one-branch phase.

Now let us apply the above results to the
correlation of the electron operator given by
$V_{\v l_e}$, where $\v l_e$ is given in Eq. (\ref{l-e}).
We find that $\v l_e K^{-1} \v m_1 = 5-3(2k_1+k_2)$, which can never vanish
for
integer $k_{1,2}$. This means that the electron correlation is short ranged
in space-time in the one-branch phase. {\em It costs a finite energy to add
(remove) an electron to (from) the edge in the one-branch phase.}

We next consider a more general
$n$-electron operators described by
$\v l_{ne}^T=( k_1, k_2, -5(n-k_1-k_2))$.
We find that $\v l_{ne} K^{-1} \v m_1 = 5n-3(2k_1+k_2)$. Thus, in the
one-branch
phase, $n$-electron operator is gapless if and only if $n$ is multiples
of $3$. The correlation of the $3m$-electron operator $V_{\v l_{(3m)e}}$ has
an algebraic decay if $ k_2=5m-2k_1$. The exponent is $5m^2$.

For a generic quasi-particle operator $V_{\v l}$, we find that
$\v l K^{-1} \v m_1 =0$ requires $l_3=2l_2-l_1$.  Only those quasi-particles
are gapless. The exponent of the quasi-particle correlation is found to be
$(2l_1+l_2)^2/5$ and the charge of the quasi-particle is $3(2l_1+l_2)/5$.

We see that the edge excitations in the one-branch phase are exactly
those of the $1/5$ Laughlin state. But the particles that form
the Laughlin state carry charge $3e$.
Such a state is described by $K=(5)$ and $\v t=(3)$.
Thus the transition from the three-branch phase to the one-branch phase
on the edge of the $\nu=9/5$ state can be viewed as a binding transition
in which electrons form triplet bound states.

To summarize, we find that when the velocity matrix $V_{ij}$ makes
$\Delta(\v
m_1)<2$ and when $\Gamma$ is large enough, the edge of the $\nu=9/5$ state
has a transition from the three-branch phase to the one-branch phase.
Charge $e$ excitations open up a finite energy gap across the transition,
while charge $3e$ excitations remain gapless in both phases.

So far, we only discussed the binding transition caused by $V_{\v m_1}$.
$V_{\v m_2}$ will cause the exactly the same binding transition due to a
symmetry between $\v m_1$ and $\v m_2$ (see \Eq{Kt} and \Eq{m12}).

To see the physical effect of the binding transition, let us consider
tunneling
between the $\nu=9/5$ and the $\nu =1$ (metal) states.\cite{C}  Before the
binding
transition, all the three branches contribute to the exponent of the
tunneling
conductance at finite temperature.  After binding transition, the first two
branches become gaped, and there is only one
gapless mode.  Furthermore, the single electron tunneling also opens up a
finite gap. Only three-electron co-tunneling is gapless, which
gives $I\propto V^{13}$ at zero temperature and $\frac{dI}{dV}|_{V=0}\propto
T^{12}$ at finite temperatures.

Experimentally, the $\nu=9/5$ state is not spin polarized.
The tunneling process that causes the
binding transition also flip spins
for single-layer
systems. Thus, strong spin-orbit coupling is necessary to see the binding
transition
in single layer systems. It might be easier to observe the binding
transition
in double-layer $\nu=9/5$ state.



\begin{references}
\bibitem{comm1}
$ \exp\left\{i(\v m_{1,2})_j \phi_j\right\}$
may carry a finite momentum, and the
factor $ e^{i  p_{1,2}x}$ is included to shift the momentum of
$ V_{\v m_{1,2}}$ to zero.
\end{references}
\end{document}